\documentclass[final,3p,times,twocolumn,number,twoside,a4paper]{elsarticle}
\usepackage{graphicx,psfrag}
\usepackage{times}
\usepackage{color}
\usepackage{amsfonts,amssymb,stmaryrd,latexsym,amsmath}

\newcommand{\beq}{\begin{equation}}
\newcommand{\eeq}{\end{equation}}
\newcommand{\Order}{\mathcal{O}}
\newcommand{\mpp}{m_{\rm p}}
\newcommand{\mn}{m_{\rm n}}
\newcommand{\mpi}{M_{\pi}}
\newcommand{\mpii}{M_{\pi^0}}

\newcommand{\Fpi}{F_\pi}

\newcommand{\eps}{\epsilon}

\renewcommand{\Re}{\text{Re}\,}

\begin{document}
\title{Precision calculation of the $\pi^-d$ scattering length and\\ its impact
on threshold $\pi N$ scattering\tnoteref{preprintno}}
\tnotetext[preprintno]{Preprint no.: FZJ-IKP-TH-2010-05, HISKP-TH-10/06}

\author[juelich,moscow]{V.\ Baru} %\ead{v.baru@fz-juelich.de}
\address[juelich]{Institut f\"{u}r Kernphysik and J\"ulich Center
        for Hadron Physics, Forschungszentrum J\"{u}lich,
        D--52425 J\"{u}lich, Germany}
\address[moscow]{Institute for Theoretical and Experimental Physics, B.\ Cheremushinskaya 25, 
        117218 Moscow, Russia}

\author[juelich,ias]{C.\ Hanhart} %\ead{c.hanhart@fz-juelich.de}
\address[ias]{Institute for Advanced Simulation,
        Forschungszentrum J\"{u}lich, D--52425 J\"{u}lich, Germany}

\author[hiskp]{M.\ Hoferichter} %\ead{hoferichter@hiskp.uni-bonn.de}
\address[hiskp]{Helmholtz-Institut f\"ur Strahlen- und
         Kernphysik and Bethe Center for Theoretical Physics, Universit\"at
        Bonn,  D--53115 Bonn, Germany}

\author[hiskp]{B.\ Kubis} %\ead{kubis@hiskp.uni-bonn.de}

\author[juelich,ias]{A.\ Nogga} %\ead{a.nogga@fz-juelich.de}

\author[hiskp,ohio]{D.\ R.\ Phillips\corref{cor1}}\ead{phillips@phy.ohiou.edu}
\address[ohio]{Institute of Nuclear and Particle Physics and Department of Physics and Astronomy, 
        Ohio University, Athens, OH 45701, USA}
\cortext[cor1]{Corresponding author}

\begin{abstract}
We present a  calculation of the $\pi^- d$ scattering
length with an accuracy of a few percent using chiral perturbation theory.
For the first time isospin-violating corrections
are included consistently.  
Using data on pionic deuterium and pionic hydrogen atoms, 
we extract  the isoscalar and isovector pion--nucleon scattering lengths
and obtain
$a^+=(7.6\pm 3.1)\cdot 10^{-3}\mpi^{-1}$ and
$a^-=(86.1\pm 0.9)\cdot 10^{-3}\mpi^{-1}$.
Via the Goldberger--Miyazawa--Oehme sum rule, this leads to a charged-pion--nucleon coupling constant  $g_c^2/4\pi = 13.69\pm0.20$.

\end{abstract}
%\pacs{13.75.Gx, 12.39.Fe, 13.40.Ks, 36.10.Gv}
\maketitle

\section{Introduction}
Hadron--hadron scattering lengths are fundamental quantities characterizing
the strong interaction, and are slowly becoming accessible to {\it ab initio}
calculations in QCD~\cite{Beane:Lattice,Torok}.  Among them, of particular
interest are pion--hadron scattering lengths: the chiral symmetry of QCD and
the Goldstone-boson nature of the pions dictate that they are
small~\cite{Weinberg66}, and their non-vanishing size is linked to fundamental
quantities like the light quark masses and condensates.  Chiral symmetry in
particular predicts that the isoscalar pion--nucleon scattering length $a^+$
is suppressed compared to its isovector counterpart $a^-$.  A precise
determination of $a^+$ would improve knowledge in many areas, e.g., dispersive
analyses of the pion--nucleon $\sigma$-term~\cite{sigmaterm}, which measures
the explicit chiral symmetry breaking in the nucleon mass due to up and down
quark masses, and is, in turn, connected to the strangeness content of the
nucleon.  But, lack of $\pi^0$ beams and neutron targets makes direct
pion--nucleon scattering experiments impossible in some charge channels,
complicating a measurement of $a^+$; the only hope for future access to the
$\pi^0 p$ scattering length lies in precision measurements of threshold
neutral-pion photoproduction~\cite{Bernstein}.  Thus, the combination of data
and theory has, until now, lacked sufficient accuracy to even establish
definitively that $a^+ \neq 0$.  $a^-$, on the other hand, serves as a vital
input to a determination of the pion--nucleon coupling constant via the
Goldberger--Miyazawa--Oehme (GMO) sum rule~\cite{GMO}. 
While the uncertainty
in $a^-$ is much less than that in $a^+$, it still contributes significantly
to the overall error bar on the sum-rule evaluation~\cite{Ericson,Metsa}.
This is one of several examples where data on pion--nucleon scattering affects
more complicated systems like the nucleon--nucleon ($NN$) interaction, and
hence has an impact on nuclear physics.

\section{Pionic atoms}
Within the last ten years new information on pion--nucleon scattering 
lengths has become available due to high-accuracy
measurements of pionic hydrogen ($\pi H$). The most recent experimental
results~\cite{Gottawidth} are
\beq
\eps_{1s}=(-7.120\pm 0.012)\,{\rm eV},~~\Gamma_{1s}=(0.823\pm 0.019)\,{\rm eV},
\label{eq:piH}
\eeq for the (attractive) shift of the $1s$ level of $\pi H$ due to strong
interactions and its width. These are connected, respectively, to the
$\pi^-$--proton scattering length, $a_{\pi^- p}$, and the charge-exchange
scattering length in the same channel~\cite{hadatoms}.  $\epsilon_{1s}$ is
related to $a_{\pi^- p}$ through an improved Deser formula~\cite{LR00} \beq
\eps_{1s}=-2\alpha^3 \mu_H^2 a_{\pi^-p}(1+K_\eps+\delta_\eps^{\rm vac}),
\label{eq:eps1s}
\eeq
where 
$\alpha=e^2/4\pi$, $\mu_H$ is the reduced mass of $\pi H$, $K_\eps=2\alpha(1-\log \alpha)\mu_Ha_{\pi^-p}$, and 
$\delta_{\eps}^{\rm vac}=2\delta \Psi_H(0)/\Psi_H(0)=0.48\%$ is the effect of 
vacuum polarization on the wave function at the origin~\cite{vac_pol}. Further, 
the width is given by~\cite{zemp}
\beq
\Gamma_{1s}=4\alpha^3\mu_H^2p_1\Big(1+\frac{1}{P}\Big)\big(a_{\pi^-p}^{\rm cex}\big)^2\big(1+K_\Gamma+\delta_\eps^{\rm vac}\big),
\label{eq:Gamma1s}
\eeq
with 
\begin{align}
K_\Gamma &=4\alpha(1-\log\alpha) \mu_Ha_{\pi^-p} \nonumber\\
& +2\mu_H(\mpp+\mpi-\mn-\mpii)(a_{\pi^0n})^2.
\end{align}  
Here $\mpp$, $\mn$, $\mpi$, and $\mpii$ are the masses of the proton, the neutron, and the charged and neutral pions, respectively,
$p_1$ is the momentum of the outgoing $n \pi^0$ pair, and the Panofsky ratio~\cite{Panofsky}
\beq
P=\frac{\sigma(\pi^- p\rightarrow \pi^0 n )}{\sigma(\pi^- p\rightarrow n \gamma )}=1.546\pm 0.009
\eeq
incorporates the effect due to the radiative decay channel of $\pi H$.
The pertinent scattering lengths 
are related to $a^\pm$ via~\cite{HKM} 
\beq
a_{\pi^-p}=\tilde a^+ +a^- + \Delta \tilde a_{\pi^- p}, ~~ a_{\pi^-p}^{\rm cex}=-\sqrt{2}\,a^-+\Delta a_{\pi^-p}^{\rm cex}.
\label{eq:IVrelns}
\eeq
Throughout we follow the notation of Ref.~\cite{HKM} for the different $\pi N$ channels, and have $\tilde{a}^+$ as $a^+$ plus a fixed shift explained below (see Sect.~\ref{sec:twobody}).
The other shifts in Eq.~\eqref{eq:IVrelns} take values $\Delta \tilde a_{\pi^- p}=(-2.0\pm 1.3)\cdot 10^{-3}\mpi^{-1}$, and 
$\Delta a_{\pi^- p}^{\rm cex}=(0.4\pm 0.9)\cdot 10^{-3}\mpi^{-1}$~\cite{HKM}. This
accounts for isospin-violating effects up to next-to-leading order (NLO) in the chiral expansion.

Equations~\eqref{eq:eps1s}, \eqref{eq:Gamma1s}, and \eqref{eq:IVrelns} permit
an extraction of $a^-$ and $\tilde a^+$. However, further experimental
information leads to better control of systematics and could enhance the
accuracy of the scattering-length determination. Consequently, additional
measurements of pion--nucleus atoms are of high interest---especially for
atoms with isoscalar nuclei, as they provide better access to $a^+$. Here we
use state-of-the-art theory to perform a combined analysis of the recent data
for pionic deuterium ($\pi D$) as well as the numbers in Eq.~\eqref{eq:piH}
for $\pi H$. The resulting values for $a^-$ and $a^+$ are of unprecedented
accuracy.

In this work we focus on the strong shift, $\eps_{1s}^D$, of the $1s$ level of pionic
deuterium, which is related to the real part of the $\pi^-$--deuteron scattering length, $\Re a_{\pi^- d}$, by an improved Deser formula analogous to Eq.~(\ref{eq:eps1s})~\cite{mrr}:
\beq
\eps_{1s}^D=-2\alpha^3\mu_D^2\Re a_{\pi^- d}\left(1+K_D+\delta_{\eps^D}^{\rm vac}\right).
\label{eq:deutDeser}
\eeq
In Eq.~(\ref{eq:deutDeser}) we have $\delta_{\eps^D}^{\rm vac}=2\delta \Psi_D(0)/\Psi_D(0)=0.51\%$ \cite{vac_pol}, $K_D=2\alpha(1-\log \alpha)\mu_D \Re a_{\pi^- d}$, and $\mu_D$ as the $\pi D$ reduced mass.

\section{The pion--deuteron scattering length}

The real part of $a_{\pi^- d}$ can be decomposed into its two- and three-body contributions as:
\begin{equation}
\Re a_{\pi^- d}= a_{\pi^- d}^{(2)}+ a_{\pi^- d}^{(3)}.
\end{equation}
It is in $a_{\pi^- d}^{(2)}$ that $a^+$ resides. Therefore, $a_{\pi^-
 d}^{(3)}$ must be calculated reliably if measurements of $\eps_{1s}^D$ are
going to be profitably exploited to get information on $a^+$.

Thus, the bulk of the rest of this paper describes a calculation of $a_{\pi^-
 d}^{(3)}$ in chiral perturbation theory ($\chi$PT). This quantity can be expressed as
\begin{equation}
a_{\pi^- d}^{(3)}=a^{\rm str}+a^{{\rm disp}+\Delta}+a^{\rm EM},
\label{eq:apid3}
\end{equation}
where $a^{\rm str}$ defines the strong contribution,  $a^{{\rm disp} + \Delta}$ involves two-nucleon and $\Delta$-isobar--nucleon 
intermediate states, as well as diagrams with crossed pion lines, and $a^{\rm EM}$ involves photon-exchange contributions. This last piece is present because isospin
violation from the up-down quark mass difference and electromagnetic effects
must be taken into account
(as in  
Ref.~\cite{HKM} 
we use a counting where $e \sim p$). 
Consistent consideration
of such effects is a key
advance made in this paper. We now deal with each of the contributions in
Eq.~\eqref{eq:apid3}, before returning to $a_{\pi^- d}^{(2)}$.

\subsection{Strong contributions ($a^{\rm str}$)}
\begin{figure}
\centering
\includegraphics[width=\linewidth]{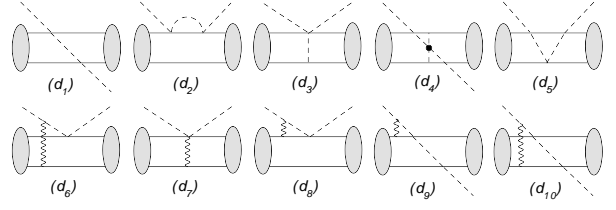}
\caption{Topologies for $\pi^- d$ scattering. 
Solid, dashed, and wiggly lines denote nucleons, pions, and photons, respectively. 
The blobs indicate  the deuteron wave functions.}
\label{fig:Feynman}
\end{figure}
The leading diagrams contributing to $a^{\rm str}$ are shown in the first line
of Fig.~\ref{fig:Feynman}. So far no counting scheme is known that permits consistent,
realistic, and simultaneous consideration of the two- and three-body operators which contribute to 
$\pi^-d$ scattering.
However, each  of these operators can be calculated independently, i.e.\ within its class,  with a controlled uncertainty. 
In particular, Ref.~\cite{Liebig:2010ki} showed how the original counting by
Weinberg~\cite{weinberg,beane98} can be modified such that the three-body contributions
to $a_{\pi^- d}$ are calculated to very high accuracy. 
Since isospin breaking in the two-body sector is also well under control~\cite{HKM}, this permits
a precise extraction of $\tilde a^+$.
Therefore, we now discuss the power counting
for all contributions to $a^{\rm str}$ relative to the leading, $\Order(1)$, diagram $(d_1)$.  

In this counting there is a $(N^\dagger N)^2\pi^\dagger\pi$ contact
term associated with the short-distance pieces of the integrals, which enters
with an unknown coefficient at $\Order(p^2)$. This contribution cannot easily
be determined from data, and is a key source of uncertainty in our result.
With $p \sim M_\pi/\mpp$, we anticipate an accuracy of a few per cent for
threshold $\pi^- d$ scattering. This expectation is substantiated by the
sensitivity of our integrals to the choice of the deuteron wave function (see
below). There we see a residual scale dependence of about $5\%$: an
independent estimate of the contact term's effect.  

But, to reach this
accuracy, we must include all three-body terms up to $\Order(p^{3/2})$.
In Ref.~\cite{beane} it was shown that the sum of all NLO, $\Order(p)$, contributions
vanishes in the isospin limit, corrections to which only enter at
$\Order(p^3)$. Thus, the diagrams we need to consider up to $\Order(p)$ are
$(d_1)$--$(d_4)$ in Fig.~\ref{fig:Feynman}.  Note that although we count
$(d_5)$ as $\Order(p^2)$, its value is enhanced by a factor of $\pi^2$ due to its
topology of two successive Coulombic
propagators~\cite{Liebig:2010ki,beane}.
Similar enhancements are present for
all terms of the multiple-scattering series. Despite this, the
multiple-scattering series converges quite quickly: we find from an explicit
calculation that the sum of the first two terms $(d_1)$ and $(d_5)$ differs
from the full result by only $0.1\cdot 10^{-3}\mpi^{-1}$.  Note that the next diagram,
where the pion leaves the two-nucleon
system after four $\pi N$ interactions on alternating nucleons, is
logarithmically divergent, and therefore seems to necessitate a contact term. 
As the terms in the multiple-scattering series are enhanced as just
described, we expect this contact term to also be enhanced.
However, that enhancement is not enough to overcome the $p^4$ suppression relative to the 
leading, double-scattering, piece of $a_{\pi^- d}$,
and so any such contact term has an appreciably smaller
effect than the $\Order(p^2)$ contact term.
Therefore its contribution
does not impact the uncertainty estimate given above. 

To achieve the requisite accuracy for our $\tilde{a}^+$ extraction we also need to include isospin-violating corrections from the different masses of the
proton and neutron and charged and neutral pions in the
diagrams $(d_1)$--$(d_4)$. 
We then express the sum of diagrams in the first row of Fig.~\ref{fig:Feynman} as:
\begin{equation}
a^{\rm str}=a^{\rm static} + a^{\rm static}_{\rm NLO}
+ a^{\rm cut}+ \Delta a^{(2)} 
+a^{\pi \pi}+a^{\rm {triple}}.
\label{eq:astr}
\end{equation}
The first four terms arise from diagrams $(d_1)$ and  $(d_2)$. However,
$(d_2)$ is partly accounted for in the two-body contribution $a_{\pi^-
 d}^{(2)}$. In
order to treat the three-body dynamics properly we must replace the
contribution of the two-body ($\pi N$) cut there by that of the three-body
($\pi NN$) cut~\cite{recoil}. The necessary integrals can be rearranged as in
Eq.~\eqref{eq:astr} (for details see \cite{longJOB}).
$a^{\rm static}$ corresponds to $(d_1)$ evaluated with a static pion
propagator, and is numerically by far the dominant contribution. $a^{\rm
 static}_{\rm NLO}$ incorporates recoil corrections to the static pion
propagator; $a^{\rm cut}$ comprises effects due to the three-body $\pi^0 nn$
and $\pi^- pn$ cuts, and $\Delta a^{(2)}$ emerges as an isospin-violating
correction in this rearrangement.  (In principle, there are also contributions
with $P$-wave interactions between nucleons in the intermediate state, but
they are of higher order.)  Finally, $a^{\pi \pi}$ in Eq.~\eqref{eq:astr} is
determined by $(d_3)$ and $(d_4)$, while $a^{\rm triple}$ results from
$(d_5)$. Isospin-breaking corrections to the $\pi N$ scattering lengths that
appear in $a^{\rm str}$ are relevant only for $a^{\rm static}$, to which they
contribute about $1\%$.

Our power counting is based on dimensional analysis assuming all integrals
scale only with $M_\pi$. In fact, the integrals in Eq.~\eqref{eq:astr} involve
other scales too: $\sqrt{M_\pi \epsilon}$---due to the three-body cut---and
$\sqrt{\mpp \epsilon}$, thanks to the deuterium wave functions ($\epsilon$
is the deuteron binding energy). At first glance, the presence of a three-body cut in the integral for $a^{\rm cut}$
makes it
appear to be enhanced over its naive $\chi$PT order by $\sqrt{\mpp/M_\pi}$~\cite{gammad}. However, this turns out not to be the case, because the Pauli principle and the spin-isospin character of the leading $\pi
N$ scattering operator ensure that the intermediate $NN$ state in $(d_1)+(d_2)$ is projected
onto a $P$-wave~\cite{recoil}. In consequence the scales $\sqrt{M_\pi
 \epsilon}$ and $\sqrt{\mpp \epsilon}$ do not enter the final result: any
enhanced contribution cancels due to a subtle interplay between the two diagrams that is dictated by the Pauli principle. The combined integral is,
as originally assumed in establishing the $\chi$PT ordering of diagrams, then dominated by
momenta of order $M_\pi$.

The results for the pieces of $a^{\rm str}$ are given in 
Table~\ref{table:strong}. They produce a total:
\beq  a^{\rm str} = (-22.6\pm 1.1\pm 0.4)\cdot
10^{-3}\mpi^{-1}. \eeq 
The first error comes from the evaluation of
all mentioned diagrams using different deuteron wave functions (we use NNLO chiral (five  wave functions with different cutoffs)~\cite{NNLO}, CD Bonn~\cite{CDBonn}, and AV 18~\cite{AV18} potentials), while the second is
due to the uncertainty in the isospin-breaking shifts in the $\pi N$ scattering lengths~\cite{HKM}.

\begin{table}
\centering
\caption{Strong contributions to $a^{(3)}_{\pi^- d}$ in units of $10^{-3} \mpi^{-1}$. Here and below results are quoted
for $a^-=86.1\cdot 10^{-3}\mpi^{-1}$. For the band in
Fig.~\ref{fig:bands} the full $a^-$ dependence is taken
into account.}
\label{table:strong}
\begin{tabular}{cccc}
\hline\hline
$a^{\rm static}$ & $-24.1\pm 0.7$ & $a^{\rm static}_{\rm NLO}$ & $3.8\pm 0.2$\\\hline
$a^{\rm cut}$ & $-4.8\pm 0.5$ & $a^{\rm triple}$  & $2.6\pm 0.5$ \\\hline 
$a^{\pi\pi}$ & $-0.2\pm 0.3$ & $\Delta a^{(2)}$ & $0.2$\\
\hline\hline
\end{tabular}
\end{table}

\subsection{Photon loops ({$a^{\rm EM}$})}

Effects in this class due to photons with momenta of order $\alpha M_\pi$ are included in observables via the improved 
Deser formula. Thus, our calculation of $a_{\pi^- d}^{(3)}$ should include contributions from momenta above
$\alpha M_\pi$. The leading contributions due to the exchange of (Coulomb)
photons of momenta of order $M_\pi$ between the $\pi^-$ and
the proton  are shown in the second row of Fig.~\ref{fig:Feynman}: $(d_6)$, $(d_7)$, and $(d_8)$.  
Photon exchange is perturbative at $|{\bf k}| \sim M_\pi$ (in contrast to the hadronic-atom regime where the photon ladder needs to be resummed), and the pertinent pieces of these graphs enter at $\Order(p)$ relative to $(d_1)$. 
Such effects in the other diagrams are of a
higher $\chi$PT order than we are considering here. 

However, diagrams $(d_6)$ and
$(d_8)$--$(d_{10})$ are reducible in the sense originally defined by
Weinberg~\cite{weinberg}, with the $\pi NN$ intermediate state involving
relative momenta of order $\sqrt{M_\pi \epsilon} \ll M_\pi$. Furthermore, in these
diagrams, this state can occur with the $NN$ pair in an $S$-wave, so we must
also allow for the possibility of $NN$ interactions while the pion is ``in
flight''. When this is done we see that these four diagrams have an infrared
divergence in the limit $\epsilon \rightarrow 0$, being enhanced by
$\sqrt{M_\pi/\epsilon}$ as compared to their naive $\chi$PT order.  

In order to avoid double counting we must also subtract from the resulting expressions for $(d_6)$ and 
$(d_8)$--$(d_{10})$ (plus $NN$ intermediate-state interactions) the quantum-mechanical
interference between a zero-range (strong) pion--deuteron potential, proportional to $a_{\pi^- d}$, and
the Coulomb interaction. That interference is already accounted for 
in the improved Deser formula~\eqref{eq:deutDeser}. Note though, that Eq.~(\ref{eq:deutDeser}) only accounts for intermediate-state pion
(and deuteron) momenta of order $\alpha M_\pi$. In particular, deuteron structure plays no role in its derivation. 

After the pieces of  $(d_6)$ and 
$(d_8)$--$(d_{10})$ that are already included in Eq.~\eqref{eq:deutDeser} are removed the result is finite. 
The remaining, finite parts of $(d_6)$ and 
$(d_8)$--$(d_{10})$ capture the effects of momenta $\gg \alpha M_\pi$ in these loops. These contributions are defined here to be part of
$a_{\pi^- d}$, and must be calculated explicitly. In particular, they include effects in the loop which arise from the electromagnetic and pion--deuteron ``form factors'': the manner
in which the finite extent of the deuteron modifies the loop integral for momenta well above the hadronic atom scale $\alpha M_\pi$~\cite{longJOB}. 

This contribution to $a_{\pi^- d}$ is ostensibly large,
since it is an infrared-sensitive integral that potentially has contributions from momenta of order $\sqrt{M_\pi \epsilon}$. 
But, analysis
analogous to Ref.~\cite{recoil_kd} shows that this particular piece of the integral is zero because of symmetry arguments. When the $NN$ pair is in an $S$-wave it
can be written as a sum of overlaps between $NN$ wave functions in the continuum and the deuteron bound state, 
and orthogonality then guarantees that the result is zero. 
In the case of an intermediate $NN$ $P$-wave pair it is the
Pauli principle that causes the cancellation~\cite{longJOB}.

There is still a possible contribution in the loop from momenta of order $\sqrt{\mpp
 \epsilon}$. This would be enhanced by $M_\pi/\sqrt{\mpp \epsilon}$ compared
to its naive $\chi$PT order, and so could be relevant for our analysis. Direct
evaluation of this part of $(d_6)$ and $(d_8)$---including the diagrams with
$NN$ interactions in an $S$-wave---yields a contribution to $a_{\pi^- d}$ of
$-0.04\, \tilde{a}^+$. (Isospin-breaking shifts of $\tilde{a}^+$
can be added here, but do not change the prefactor.) Replacing the single
$\pi N$ scattering of these diagrams by double scattering as in $(d_9)$ and
$(d_{10})$ gives effects larger by a factor of $a_{\pi^- d}/
2\tilde{a}^+$, but, despite their being infrared enhanced, the impact of such pieces on $a_{\pi^- d}$ is still significantly less
than our theoretical uncertainty.

This leaves us needing to consider only effects from momenta $M_\pi$ in
diagrams $(d_6)$--$(d_8)$. As with $(d_2)$ in $a^{\rm str}$, parts of these
diagrams are already included in $a_{\pi^- d}^{(2)}$, but this can be dealt
with along the same lines~\cite{longJOB}.  The result is: 
\beq a^{\rm EM} = (0.94\pm
0.01)\cdot 10^{-3}\mpi^{-1}, \eeq
where the error again reflects the
wave-function dependence. Thus, virtual photons with $|{\bf k}| \sim M_\pi$
increase $\Re a_{\pi^- d}$ by about 4\%.

\subsection{Dispersive and Delta(1232) corrections ($a^{{\rm disp}+\Delta}$)}
These  produce effects in $a_{\pi^- d}$ that scale with half-integer powers of $p$~\cite{disp,delta}. 
Their leading contribution is $\Order(p^{3/2})$ relative to $(d_1)$, and is computed here using a calculation for $NN\to d\pi$
up to NLO in $\chi$PT~\cite{pp2dpi}. 
Note that although we include Delta(1232) effects in the $\pi NN\to \pi NN$
transition operator, it is not necessary to account for the Delta(1232) as an explicit degree of freedom when computing
the deuteron wave function. Its effects in the $NN$
potential at energies of order $\epsilon$ enter only at relative $\Order(p^2)$~\cite{delta}.
In Refs.~\cite{disp,delta} all integrals were cut off  at
$1\,{\rm GeV}$; we have checked that this does not introduce
additional uncertainty and obtain:
\begin{equation}
a^{{\rm disp}+\Delta}=(-0.6\pm 1.5)\cdot 10^{-3}\mpi^{-1}.
\end{equation}
Since this is at the limit of our desired accuracy we need not include isospin-violating 
corrections to $a^{{\rm disp}+\Delta}$.

\subsection{The two-body part ($a^{(2)}_{\pi^-d}$)}
\label{sec:twobody}
As alluded to above, it is not possible to isolate $a^+$ in analyses of $\pi H$ and $\pi D$.
Information on the isoscalar scattering length can only be extracted as a combination $\tilde{a}^+$, in which the low-energy constants $c_1$ (which occurs because its impact on $a^+$ is proportional to the neutral-pion mass squared) and $f_1$ (which denotes the leading 
isoscalar electromagnetic correction) also appear~\cite{mrr}:
\begin{equation}
\tilde a^+ \equiv a^+ + \frac{1}{1+\mpi/\mpp}
\bigg\{\frac{\mpi^2-\mpii^2}{\pi \Fpi^2}c_1-2\alpha f_1\bigg\}.
\end{equation}

In the two-body part of $a_{\pi^- d}$, $\tilde{a}^+$ is further shifted, as shown in the NLO analysis of Ref.~\cite{HKM}:
\begin{align}  
a_{\pi^- d}^{(2)}&=
\frac{2 \mu_D}{\mu_H} \left(\tilde a^++\Delta \tilde a^+\right), \notag\\
\Delta \tilde a^+&=(-3.3\pm 0.3)\cdot 10^{-3}\mpi^{-1}\label{aplustilde}.
\end{align}

\begin{figure}
\centering
\includegraphics[width=\linewidth,clip]{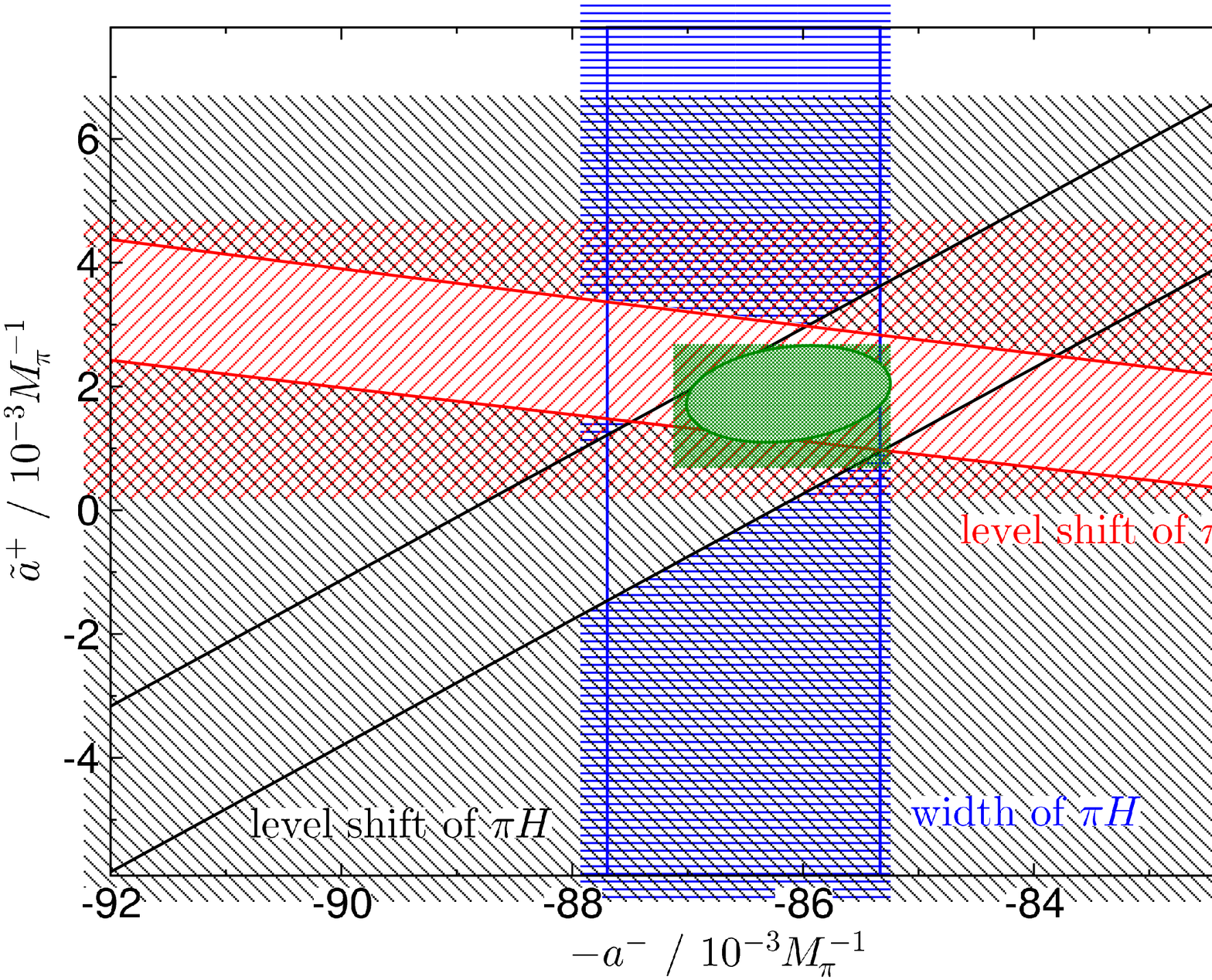}
\caption{Combined constraints in the $\tilde a^+$--$a^-$ plane from data on the width and energy
shift of $\pi H$, as well as the $\pi D$ energy shift.}
\label{fig:bands}
\end{figure}

\section{Results and Discussion}

We now add together all the individual 
contributions. Amusingly, most of the additional three-body corrections considered in this study
accidentally cancel: 
$\Delta a^{(2)} + a^{\rm static}_{\rm NLO}
+ a^{\rm cut} + a^{\rm EM} = (0.1\pm 0.7)\cdot 10^{-3} \mpi^{-1}$.
For this reason, the main impact of our analysis on the extraction of pion--nucleon scattering lengths turns out to be due
to the NLO isospin-breaking corrections in the two-body part~\cite{Hoferichter:CD09}.

The energy shift of $\pi D$ has recently  been 
remeasured as~\cite{Gottapid}
\beq
\eps_{1s}^D=(2.356\pm 0.031)\,{\rm eV}.
\eeq
Combining this result, the dependence of the $\pi^- d$ scattering length on $\tilde a^+$ and
$a^-$, and the results for $\pi H$ discussed above, we find the constraints depicted in 
Fig.~\ref{fig:bands}. The combined $1\sigma$ error ellipse yields 
\beq
\tilde{a}^+=(1.9\pm 0.8)\cdot 10^{-3} \mpi^{-1},~~ a^-=(86.1\pm 0.9)\cdot 10^{-3}\mpi^{-1},
\label{eq:atilde+}
\eeq
with a correlation coefficient $\rho_{a^-\tilde a^+}=-0.21$.  
We find that the inclusion of the $\pi D$ energy shift reduces the uncertainty of $\tilde{a}^+$
by more than a factor of 2.
Note that in the case of the $\pi H$ level shift the width of the band is dominated
by the theoretical uncertainty in $\Delta \tilde a_{\pi^- p}$,
whereas for the $\pi H$ width the experimental error is about $50\%$ larger than the theoretical one.
The uncertainty in $a^{{\rm disp} + \Delta}$ is the largest contribution to the $\pi D$ error band, see Table~\ref{table:piDlevel}.
The wave-function averages contribute about $0.5 \cdot 10^{-3} M_\pi^{-1}$ to the overall uncertainty in $\tilde a^+$, 
which is in line with the estimated impact on $a_{\pi^- d}$ of the $\Order(p^2)$---relative
to $(d_1)$---contact term.

\begin{table}
\centering
\caption{Individual contributions to the error on $\tilde{a}^+$ are added in quadrature to obtain the uncertainty depicted in the bands of Fig.~\ref{fig:bands}. Each row below gives the impact of one source of error as a percentage of that total. The first row is the impact of the experimental uncertainty in $\epsilon_{1s}^D$, the second gives the uncertainty in the isospin-breaking shifts of $\pi N$ scattering lengths that occur in $a^{\rm str}$, and the third row is the uncertainty in $\Delta\tilde{a}^+$ according to Eq.~\eqref{aplustilde}. The final two rows show the impact of uncertainties in our calculation of $\Re a_{\pi^- d}^{(3)}$, as described in the text.}
\label{table:piDlevel}
\vspace*{0.1in}
\begin{tabular}{cc}
\hline\hline
		 $\epsilon_{1s}^D$ & 16\%\\ \hline
		 $\Delta \tilde a_{\pi^-p}, \Delta a_{\pi^- p}^{\rm cex}$ & 21 \% \\\hline
		   $\Delta \tilde{a}^+$ & 30 \%\\ \hline
		 $a^{{\rm disp}+\Delta}$ & 75\% \\ \hline
Wave-function averages & 53\%\\
\hline\hline
\end{tabular}
\end{table}

Taken together with $c_1=(-1.0\pm 0.3)\,{\rm GeV}^{-1}$~\cite{longJOB} 
and the rough estimate $|f_1|\leq 1.4 \,{\rm GeV}^{-1}$~\cite{f1},
Eq.~\eqref{eq:atilde+} yields a non-zero $a^+$ at better than the $95\%$ confidence level:
\beq
a^+=(7.6\pm 3.1)\cdot 10^{-3}\mpi^{-1}.
\eeq

The final result for $a^+$ is only a little larger than several of the contributions considered in our analysis. 
This emphasizes the importance of a systematic ordering scheme, and a careful treatment of
isospin violation and three-body dynamics. 
A reduction of the theoretical uncertainty beyond that of the present analysis will be hard to achieve without 
additional QCD input that helps pin down the unknown contact-term contributions 
in both the $\pi N$ and $\pi NN$ sectors. 

Finally, these results allow us to infer the charged-pion--nucleon coupling
constant, $g_c$, from the GMO sum rule, with isospin-violating corrections
to the $\pi N$ scattering lengths
fully under control for the first time.  Inspired by Ref.~\cite{Ericson}, we
take $a_{\pi^-p}$ extracted from Eq.~\eqref{eq:eps1s}, $a_{\pi^-p}+a_{\pi^-n}$
from our $a_{\pi^-d}$ analysis, and $a_{\pi^- n}-a_{\pi^+p}$ from
Ref.~\cite{HKM}, yielding $g_c^2/4\pi=13.69\pm 0.12\pm
0.15$~\cite{longJOB}. (Here the first error is due to the scattering lengths
and the second to an integral over $\pi^\pm p$ cross
sections~\cite{Ericson,Metsa}.) This is in agreement with determinations from
$NN$~\cite{deSwart} and $\pi N$~\cite{arndt} scattering data.

\section*{Acknowledgments} 

We  thank D.~Gotta, A.~Kudryavtsev, U.-G.~Mei\ss ner, A.~Rusetsky, M.~Sainio, and A.~W.~Thomas for useful discussions.
This research was supported by the DFG (SFB/TR 16, ``Subnuclear Structure of Matter''), DFG-RFBR grant (436 RUS 113/991/0-1), the Mercator Programme of the DFG, the Helmholtz Association through funds provided
to the virtual institute ``Spin and strong QCD'' (VH-VI-231) and the young investigator group ``Few-Nucleon Systems in Chiral Effective Field Theory'' (grant VH-NG-222), the Bonn-Cologne Graduate School of Physics and Astronomy, 
the project ``Study of Strongly Interacting Matter''
(HadronPhysics2, grant No.~227431) under the 7th Framework Programme of the EU, 
the US Department of Energy (Office of Nuclear Physics, under contract No.~DE-FG02-93ER40756 with Ohio University), and the Federal Agency of Atomic Research of the Russian Federation (``Rosatom'').
Computing resources were provided by the JSC, J\"ulich, Germany.

\end{document}